\documentclass{emulateapj}
\usepackage{graphicx}
\usepackage{txfonts}
\usepackage{natbib}
\usepackage[scaled]{helvet}
\usepackage{epsfig}
\usepackage{url}
\usepackage[normalem]{ulem}
\usepackage{color}
\bibpunct{(}{)}{;}{a}{}{,}
\newcommand{\bjdtdb}{\ensuremath{\rm {BJD_{TDB}}}}

\newcommand{\teff}{\ensuremath{T_{\rm eff}}}

\newcommand{\ecosw}{\ensuremath{e\cos{\omega_*}}}
\newcommand{\esinw}{\ensuremath{e\sin{\omega_*}}}
\newcommand{\msun}{\ensuremath{\,M_\Sun}\space}
\newcommand{\rsun}{\ensuremath{\,R_\Sun}\space}

\newcommand{\kms}{\,km\,s$^{-1}$}
\newcommand{\minus}{\scalebox{0.75}[1.0]{$-$}}

\begin{document}

\title{\bf A Bright Short Period M-M Eclipsing Binary from the KELT Survey: Magnetic Activity and the Mass-Radius Relationship for M-dwarfs}

\author{
Jack B. Lubin$^1$,
Joseph E. Rodriguez$^{2,\star}$}\thanks{$^{\star}$Corresponding Author, E-mail: joseph.rodriguez@cfa.harvard.edu} 
\author{George Zhou$^2$, 
Kyle E. Conroy$^{1}$,
Keivan G. Stassun$^{1,3}$, 
Karen Collins$^{1,3}$, 
Daniel J. Stevens$^{4}$,
Jonathan Labadie-Bartz$^{5}$, 
Christopher Stockdale$^{6,7}$, 
Gordon Myers$^{6,8}$, 
Knicole D. Col\'on$^{9}$, 
Joao Bento$^{10}$, 
Petri Kehusmaa$^{11}$,
Romina Petrucci$^{12,13}$,
Emiliano Jofr\'e$^{12,13}$,
Samuel N. Quinn$^{2}$,
Michael B. Lund$^{1}$, 
Rudolf B. Kuhn$^{14}$, 
Robert J. Siverd$^{15}$, 
Thomas G. Beatty$^{16,17}$, 
Caisey Harlingten$^{11}$,
Joshua Pepper$^{5}$, 
B. Scott Gaudi$^{4}$, 
David James$^{18}$, 
Eric L. N. Jensen$^{19}$,
Daniel Reichart$^{20}$,
Lucyna Kedziora-Chudczer$^{21,22}$, 
Jeremy Bailey$^{21,22}$, and 
Graeme Melville$^{21,22}$ 
}


\affil{$^1$Department of Physics and Astronomy, Vanderbilt University, 6301 Stevenson Center, Nashville, TN 37235, USA}
\affil{$^2$Harvard-Smithsonian Center for Astrophysics, 60 Garden Street, Cambridge, MA 02138, USA}
\affil{$^3$Department of Physics, Fisk University, 1000 17th Avenue North, Nashville, TN 37208, USA}
\affil{$^{4}$Department of Astronomy, The Ohio State University, Columbus, OH 43210, USA}
\affil{$^5$Department of Physics, Lehigh University, 16 Memorial Drive East, Bethlehem, PA 18015, USA}
\affil{$^{6}$American Association of Variable Star Observers, 49 Bay State Rd., Cambridge, MA 02138, USA}
\affil{$^{7}$Hazelwood Observatory, Churchill, Australia}
\affil{$^{8}$5 Inverness Way, Hillsborough, CA 94010, USA}
\affil{$^{9}$NASA Goddard Space Flight Center, Greenbelt, MD 20771, USA}
\affil{$^{10}$Research School of Astronomy and Astrophysics, Australian National University, Canberra, ACT 2611, Australia}
\affil{$^{11}$Searchlight Observatory Network, Harlingten San Pedro de Atacama Observatory}
\affil{$^{12}$Observatorio Astron\'omico de C\'ordoba (OAC), Laprida 854, X5000BGR, C\'ordoba, Argentina}
\affil{$^{13}$Consejo Nacional de Investigaciones Cient\'ificas y T\'ecnicas (CONICET), Argentina}
\affil{$^{14}$South African Astronomical Observatory, P.O. Box 9, Observatory 7935, South Africa}
\affil{$^{15}$Las Cumbres Observatory Global Telescope Network, 6740 Cortona Drive, Suite 102, Santa Barbara, CA 93117, USA}
\affil{$^{16}$Department of Astronomy \& Astrophysics, The Pennsylvania State University, 525 Davey Lab, University Park, PA 16802}
\affil{$^{17}$Center for Exoplanets and Habitable Worlds, The Pennsylvania State University, 525 Davey Lab, University Park, PA 16802}
\affil{$^{18}$Astronomy Department, University of Washington, Box 351580, Seattle, WA 98195, USA}
\affil{$^{19}$Department of Physics and Astronomy, Swarthmore College, Swarthmore, PA 19081, USA}
\affil{$^{20}$Department of Physics and Astronomy, University of North Carolina at Chapel Hill, Chapel Hill, NC 27599-3255, USA}
\affil{$^{21}$School of Physics, University of New South Wales, Sydney, NSW 2052, Australia}
\affil{$^{22}$Australian Centre for Astrobiology, University of New South Wales, Sydney, NSW 2052, Australia}
\shorttitle{M-M EB}

\begin{abstract}

We report the discovery of KELT J041621-620046, a moderately bright (J$\sim$10.2) M dwarf eclipsing binary system at a distance of 39$\pm$3 pc. KELT J041621-620046 was first identified as an eclipsing binary using observations from the Kilodegree Extremely Little Telescope (KELT) survey. The system has a short orbital period of $\sim$1.11 days and consists of components with M$_1$ = $0.447^{-0.047}_{+0.052}\,M_\odot$ and M$_2$ = $0.399^{-0.042}_{+0.046}\,M_\odot$ in nearly circular orbits. The radii of the two stars are R$_1$ = $0.540^{-0.032}_{+0.034}\,R_\odot$ and R$_2$ = $0.453\pm0.017\,R_\odot$. Full system and orbital properties were determined (to $\sim$10\% error) by conducting an EBOP global modeling of the high precision photometric and spectroscopic observations obtained by the KELT Follow-up Network. Each star is larger by 17-28\% and cooler by 4-10\% than predicted by standard (non-magnetic) stellar models. Strong H$\alpha$ emission indicates chromospheric activity in both stars. The observed radii and temperature discrepancies for both components are more consistent with those predicted by empirical relations that account for convective suppression due to magnetic activity. 
\end{abstract}

\shortauthors{Lubin et al.}

\section{Introduction}

Low mass M Dwarf stars make up the majority of stars in the Milky Way, however our understanding of these systems is quite limited. Mass and radius, two of the most fundamental stellar parameters, are important to the understanding of stellar evolution. In the case of Low Mass Stars (LMSs), existing models struggle to accurately predict these basic stellar parameters, with isochrone fitting and stellar modelling of these LMSs often resulting in errors on mass and radius in excess of 10\%  \citep[e.g.][]{Torres:2002,Ribas:2006,Torres:2010,Spada:2013,Feiden:2014,Zhou:2014,Terrien:2015}. Additionally, the discovery of transiting super-Earths around M-dwarfs from MEarth \citep{Nutzman:2008, Irwin:2009} and the Kepler/K2 mission \citep{Borucki:2010, Howell:2014}, for which the estimates of planetary parameters are directly dependent on the model-determined stellar parameters, it is crucial that we better understand and model these fundamental stellar properties. One type of natural astrophysical laboratory for precisely measuring mass and radius are detached, double-lined M-dwarf eclipsing binaries (EBs). Although these systems are quite rare, they provide an opportunity to measure precise stellar parameters in a model-independent fashion. These measurements serve as direct tests of theoretical stellar models \citep[e.g.][]{Chabrier:1995, Torres:2002}.

\begin{figure*}[!ht]
  \centering\includegraphics[width=0.99\linewidth]{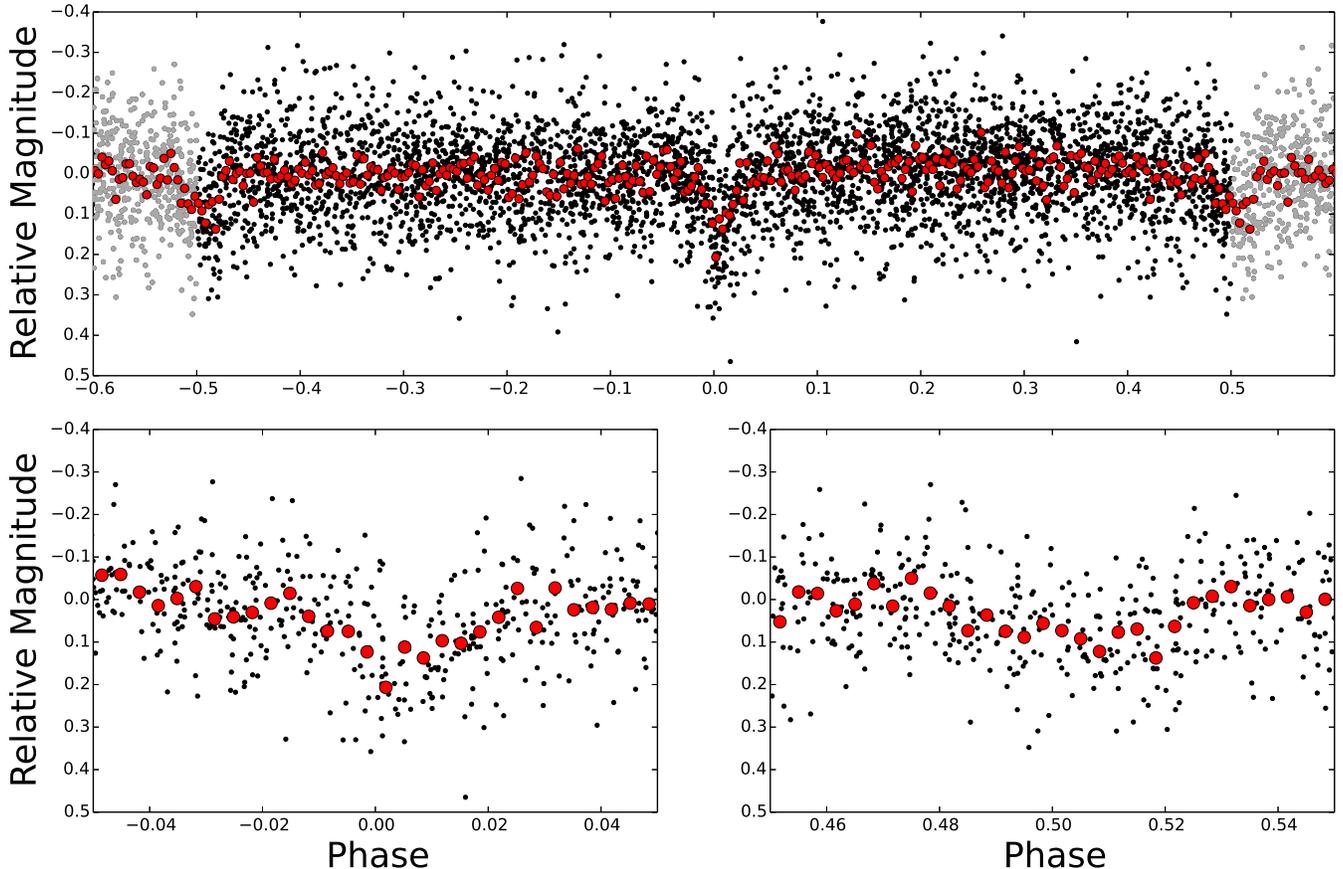}
  \caption{(Top) The KELT-South light curve of KELT J041621-620046 phase-folded to the discovery ephemeris (T$_0$ = 2455255.950528 in JD$_{TT}$, P = 1.1112884 days).  The binned data are shown in red. (Bottom) A zoom-in of the primary (left) and secondary (right) eclipses.}
\label{fig:DiscoveryLC}
\end{figure*}

Due to the intrinsic faintness of these systems, there is only a small number of M-M EBs currently known, and most of them are too faint to precisely measure the stellar parameters. One of the brightest and best-studied M-M EBs, CM Draconis (V = 12.87) \citep{Eggen:1967}, has been the target of extensive characterization allowing the measurement of the system's parameters to $\le$1\% precision \citep{Lacy:1977, Metcalfe:1996, Morales:2009}. The short orbital period ($\sim$1.27 days) suggests that the rotation period of the two stars is synchronized with the orbital period of the binary \citep{Morales:2009}. However, the stellar radii in CM Dra are 5-7\% larger than predicted by theoretical models \citep{Feiden:2014}. It has been proposed that metallicity and magnetic activity are responsible for the large radii of M dwarfs \citep{Lopez:2005, Ribas:2006, Chabrier:2007}. At $\sim$41 days, LSPM J1112+7626 is the longest period detached M-M binary discovered to date \citep{Irwin:2011}. As a result of the large semi-major axis, the stellar components in this binary have rotation periods that are not synchronized with the orbital period. Therefore, the stellar components of LSPM J1112+7626 should be more representative of field or single M-stars. However, \citet{Irwin:2011} found that the combined radii of the two stars in the LSPM J1112+7626 system are inflated by 3.8\%. Even with the discovery of additional systems, there are only a handful of M-M EBs known that can be used to test theoretical models \citep{Irwin:2011,Birkby:2012, Nefs:2013, Zhou:2015, Dittmann:2016}.



Mounting observational evidence shows that magnetically active stars are 5-15\% larger and $\sim$5\% cooler than predictions from standard stellar models \citep[see, e.g.,][]{Lopez:2007}. Empirical work suggests that magnetic activity may cause these anomalies by suppressing surface convection \citep[see, e.g.,][]{Morales:2010}. Empirical relations have been developed to correct for these effects \citep[e.g.,][]{Stassun:2012}. These corrections are important for multiple purposes, including determining the true initial mass functions of star forming regions, particularly at the low-mass end \citep[e.g.,][]{Stassun:2014}. Interestingly, from an X-ray study of nearby M-dwarfs, there is no observed difference in the X-ray behavior between single and binary systems \citep{James:2000}.  

In this paper, we present the discovery of KELT J041621-620046, a bright (J$\sim$10.2) 0.45+0.40 \msun eclipsing binary. This system is one of the brightest known detached M-M EB, with a very short orbital period of 1.11 days. With only about three dozen M-M EBs known, most with $V>$14, KELT J041621-620046 provides the opportunity to test theoretical models for higher--mass M dwarfs.

Section \ref{sec:Observations} presents the discovery photometry and the follow-up photometric and spectroscopic observations used. Section \ref{sec:results} presents the analysis and principal results of this study, including direct measurement of the two stars' masses, radii, and other properties. In Section \ref{sec:Discussion}, we briefly discuss the results in the context of the predictions of stellar models and of empirical relations for radius and temperature anomalies in LMSs. We summarize the results in Section \ref{sec:Conclusion}.

\begin{table*}
\centering
\caption{Stellar Properties of KELT J041621-620046 obtained from the literature and from this work.}
\label{tbl:Host_Lit_Props}
\begin{tabular}{llccc}
  \hline
\hline
  Parameter & Description & Value & Source & Reference(s) \\
 Names 			& 					&2MASS J04162165-6200463& 		&			\\
			& 					& UCAC4 140-003916		& 		&			\\
			&					&WISE J041621.52-620047.1&		&			\\
			&					&				&		&			\\
$\alpha_{J2000}$	&Right Ascension (RA)& 	04:16:21.652 		& UCAC4	& \citet{Zacharias:2012, Zacharias:2013}	\\
$\delta_{J2000}$	&Declination (Dec)& 	-62:00:46.44		& UCAC4	& \citet{Zacharias:2012, Zacharias:2013}	\\
FUV         & Far UV magnitudes & 4.026$\pm$1.023 & GALEX & \citet{Gomez:2015} \\
NUV         & Near UV magnitudes & -0.002$\pm$0.111 & GALEX & \citet{Gomez:2015} \\
			&					&				&		&			\\
B\dotfill       	&UCAC4 magnitude&	15.368$\pm$0.03	& UCAC4	& \citet{Zacharias:2012, Zacharias:2013}	\\
V\dotfill			&UCAC4 magnitude& 13.877$\pm$0.01	& UCAC4	& \citet{Zacharias:2012, Zacharias:2013}	\\
R\dotfill			&UCAC4 magnitude& 13.55$\pm$0.06	& UCAC4	& \citet{Zacharias:2012, Zacharias:2013}	\\
g$^\prime$\dotfill			&UCAC4 magnitude& 14.63$\pm$0.02	& UCAC4	& \citet{Zacharias:2012, Zacharias:2013}	\\
r$^\prime$\dotfill			&UCAC4 magnitude& 13.30$\pm$0.01	& UCAC4	& \citet{Zacharias:2012, Zacharias:2013}	\\
i$^\prime$\dotfill			&UCAC4 magnitude& 12.13$\pm$0.03	& UCAC4	& \citet{Zacharias:2012, Zacharias:2013}	\\
			&					&				&		&			\\
$V$\dotfill		& APASS magnitude	& 13.892$\pm$0.071 & APASS		& \citet{Henden:2016}	\\
$B$\dotfill		& APASS magnitude	& 15.354$\pm$0.039	& APASS	& \citet{Henden:2016}	\\
			&					&				&		&			\\
J\dotfill			&2MASS magnitude& 10.234 $\pm$ 0.026	& 2MASS 	& \citet{Cutri:2003, Skrutskie:2006}	\\
H\dotfill			&2MASS magnitude& 9.589 $\pm$ 0.023	& 2MASS 	& \citet{Cutri:2003, Skrutskie:2006}	\\
K\dotfill			&2MASS magnitude& 9.391 $\pm$ 0.024	& 2MASS 	& \citet{Cutri:2003, Skrutskie:2006}	\\
			&					&				&		&			\\
\textit{WISE1}		&WISE passband& 9.239 $\pm$ 0.023	& WISE 		&\citet{Cutri:2012}	\\
\textit{WISE2}		&WISE passband&9.107 $\pm$ 0.02 & WISE 		& \citet{Cutri:2012}\\
\textit{WISE3}		&WISE passband&8.971 $\pm$ 0.02 & WISE 		& \citet{Cutri:2012}	\\
\textit{WISE4}		&WISE passband&9.239 $\pm$ 0.023& WISE 		& \citet{Cutri:2012}	\\
			&					&				&		&			\\
$\mu_{\alpha}$		& Proper Motion in RA (mas yr$^{-1}$)	&-73.8 $\pm$ 1.5& UCAC4		& \citet{Zacharias:2012, Zacharias:2013} \\
$\mu_{\delta}$		& Proper Motion in DEC (mas yr$^{-1}$)	&-70.9 $\pm$ 1.5& UCAC4		& \citet{Zacharias:2012, Zacharias:2013} \\
			&					&				&		&			\\
Distance & Distance (pc) & 39 $\pm$ 3 &  &  This work \\
    log(L$_{H\alpha}$/L$_{bol}$)$_A$ & H$\alpha$ Emission Strength for A & -3.903 $\pm$ 0.021 & & this work$^a$\\
    log(L$_{H\alpha}$/L$_{bol}$)$_B$ & H$\alpha$ Emission Strength for B & -3.839 $\pm$ 0.050 & & this work$^a$\\
\hline
\hline
\end{tabular}

\footnotesize \textbf{\textsc{NOTES}}\\
\footnotesize $^a$See \citet{Zhou:2015} for a description on how log(L$_{H\alpha}$/L$_{bol}$) was measured.\\

\end{table*}

\begin{figure*}[!ht]
  \centering\includegraphics[width=0.99\linewidth,trim = 0 0 0 0]{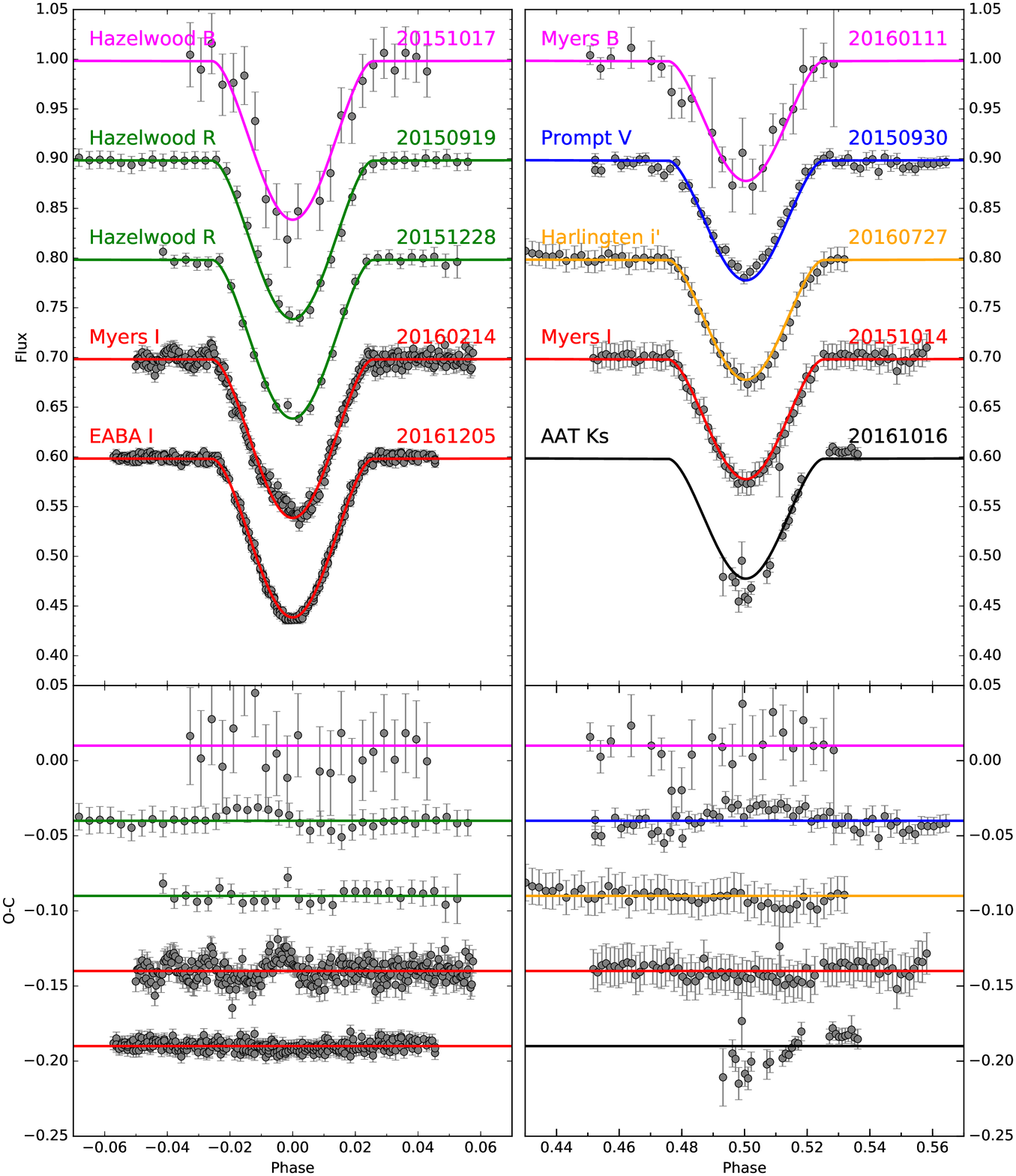}
 \vspace{-0.1in} 
  \caption{Follow-up photometric observations of the (left) primary and (right) secondary eclipses shown in phase relative to the global fit ephemeris. The best fit models are plotted over the data points. The AAT observations have been binned at step sizes of 0.001 in phase to make them comparable to the optical observations, and all other observations are plotted at their native cadences. The relative flux of each light curve has been normalized to a value of 1.0 using data in the out-of-eclipse regions, and the lower four in each panel have been offset on the flux axis for display purposes.}
    \label{fig:Followup}
\end{figure*}

\section{Observations}
\label{sec:Observations}

\subsection{KELT}
The Kilodegree Extremely Little Telescope (KELT) is an all-sky photometric survey designed to detect transiting planets around bright host stars (7$<$V$<$11). The survey uses two telescopes, KELT-North in Soniata, AZ and KELT-South at the South African Astronomical Observatory (SAAO), and each telescope uses a 4096 $\times$ 4096 pixel Apogee CCD camera with a 42 mm aperture, producing a $26^\circ$ x $26^\circ$ field of view \citep{Pepper:2007, Pepper:2012}. See \citet{Kuhn:2016} for a detailed description of the KELT-South data reduction pipeline. KELT J041621-620046 is located in KELT-South field 20, which is centered on J2000 $\alpha =$ 04$^{h}$ 36$^{m}$ 00$^{s}$ $\delta =$ -53$\degr$ 00$\arcmin$ 00$\arcsec$. 

KELT J041621-620046 itself is located at $\alpha =$ 04$^{h}$ 16$^{m}$ 21$\fs$652 $\delta =$ $\minus$62$\degr$ 00$\arcmin$ 46$\farcs$44 J2000 and was observed by KELT-South 4182 times from UT 2010 February 28 until UT 2013 April 07. The discovery KELT light curve is shown in Figure \ref{fig:DiscoveryLC}. It is also known in various catalogs by the names 2MASS J04162165-6200463, UCAC4 140-003916, and WISE J041621.52-620047.1 (See Table \ref{tbl:Host_Lit_Props}).

\subsection{Photometric Follow-up}
To better measure the KELT J041621-620046 system, we obtained higher precision photometric follow-up of the primary and secondary eclipses from the KELT Follow-Up Network (KELT-FUN). All follow-up photometry except the AAT observations were reduced using the AstroImageJ\footnote{http://www.astro.louisville.edu/software/astroimagej} package \citep{Collins:2013,Collins:2017} and the reported timestamps are $\bjdtdb$. The AAT observations were reduced following the technique described in \citet{Zhou:2014b}. As a result of the different field-of-views, pixel scales, and observing conditions for each observation, the comparison stars were selected by visually inspecting that none of them showed variability above the photometric scatter. For the AIJ aperture selection, we ran a series of reductions with a range of apertures and the optimal aperture was the one that provided the lowest RMS to a transit model. For the AAT observations, a range of apertures were used and the optimal aperture was determined by minimizing the out-of-eclipse standard deviation.  To predict both primary and secondary eclipses, we used the observing software tool TAPIR \citep{Jensen:2013}. All follow-up photometry is listed in Table \ref{tbl:detrending_parameters} and displayed in Figure \ref{fig:Followup}.

\subsubsection{Hazelwood Observatory}
Operated by Chris Stockdale, the Hazelwood Observatory is located in Victoria, Australia. This backyard observatory hosts a 0.32 m Planewave CDK12 f/8 Corrected Dall-Kirkham telescope using a SBIG ST8XME 1.5K $\times$ 1K CCD. This setup gives a 18$\arcmin$ $\times$ 12$\arcmin$ field of view with 0$\farcs$73 per pixel. Using the 0.32m telescope, a primary eclipses of KELT J041621-620046 was observed on UT 2015 September 19 in the $V$--band, UT 2015 October 17th in the $B$--band, and UT 2015 December 28th in the $R$--band. An observation of the secondary eclipse was observed on UT 2015 October 14th in the $V$--band.

\subsubsection{Myer's Observatory}
Myer's Observatory (also known as T50) is a PlaneWave Instruments CDK17 17 inch (0.43 m) f/6.8 Corrected Dall-Kirkham Astrograph telescope, located at Siding Spring, Australia. The camera is a Finger Lakes Instruments (FLI) ProLine Series PL4710 - E2V 47-10-1-353 Back Illuminated Broadband Monochrome CCD with the Basic Mid-band coating with a 15$\farcm$5 $\times$ 15$\farcm$5 field of view and a 0$\farcs$92 pixel scale. A primary eclipse of KELT J041621-620046 was observed in the $I$ filter on UT 2016 February 14. Secondary eclipses were observed in the $I$ filter on UT 2015 October 11 and in the $B$ filter on UT 2016 January 11.

\subsubsection{Skynet}
Using the Skynet network of worldwide telescopes \citep{Reichart:2005}\footnote{https://skynet.unc.edu/}, we observed a secondary eclipse of KELT J041621-620046 on UT 2015 September 30 in the $V$ band. Specifically, we used the 0.4 m Prompt5 telescope from the PROMPT (Panchromatic Robotic Optical Monitoring and Polarimetry Telescope) subset of the Skynet network located at CTIO. The Prompt5 telescope uses an Alta U47s Apogee camera with a 10\arcmin$\times$10\arcmin field of view and a 0$\farcs$59 pixel$^{-1}$ pixel scale. 

\subsubsection{Harlingten Atacama}
Located in San Pedro de Atacama, Chile, the Harlingten San Pedro de Atacama Observatory  hosts a 0.51-meter PlaneWave telescope with an Apogee Alta U42 CCD providing a 27$\arcmin$ $\times$ 27$\arcmin$ field of view and a 0$\farcs$8 pixel scale. A secondary eclipse was observed on UT 2016 July 27 in the $i^\prime$ filter.

\subsubsection{Estaci\'on Astrof\'isica de Bosque Alegre (EABA)}
The Estaci\'on Astrof\'isica de Bosque Alegre (EABA) is located in C\'ordoba, Argentina and is operated by the Observatorio Astron\'omico de C\'ordoba. The EABA hosts a 1.54-m f/4.86 telescope operated in Newtonian focus, currently equipped with a Apogee Alta U9 camera with 3070$\times$2048 9 $\mu$m-size pixels, providing a 8$\arcmin$ $\times$ 12$\arcmin$ field of view with a plate scale of 0$\farcs$25 per pixel. A full primary eclipse of KELT J041621-620046 was observed on UT 2016 December 05 in the $I$--band, adopting a 4$\times$4 binning.

\subsubsection{Anglo-Australian Telescope (AAT)}
A secondary eclipse of KELT J041621-620046 was observed on UT 2016 October 16 in the infrared $Ks$ band with the 3.9\,m Anglo-Australian Telescope IRIS2 infrared camera, located at Siding Spring Observatory. IRIS2 uses a HAWAII-1 HgCdTe $1\times1$K infrared detector, which is read out over four quadrants to provide a field of view of 7$\farcm$7 $\times$ 7$\farcm$7, resulting in a plate scale of $0\farcs4486\,\mathrm{pixel}^{-1}$. The observing strategy and data reduction procedures for the AAT-IRIS2 observations are fully described in \citet{Zhou:2014b}.

\begin{table*}
\footnotesize
 \centering
 \caption{Photometric follow-up observations and the detrending parameters found by AIJ for the global fit.}
 \label{tbl:detrending_parameters}
 \begin{tabular}{lllllllll}
    \hline
    \hline
    Observatory & Date (UT) & Filter & FOV & Pixel Scale  & Exposure (s) &  Detrending parameters \\
    \hline
    Hazelwood & UT 2015 September 19 & $V$ & 18$\arcmin$ $\times$ 12$\arcmin$ & 0.73$\arcsec$ &300 & Airmass, X(FITS) \\
    Skynet Prompt5 & UT 2015 September 30 & $V$ & 10$\arcmin$ $\times$ 10$\arcmin$  & 0.59$\arcsec$ & 300 & Airmass, Peak T1\\
    Myers & UT 2015 October 14 & $I$ &15.5$\arcmin$ $\times$ 15.5$\arcmin$ & 0.92$\arcsec$&120 & Airmass, Width T1\\
    Hazelwood & UT 2015 October 14 & $V$ & 18$\arcmin$ $\times$ 12$\arcmin$ & 0.73$\arcsec$ &300 & Time \\
    Hazelwood & UT 2015 October 17 & $B$ & 18$\arcmin$ $\times$ 12$\arcmin$ & 0.73$\arcsec$ &300 & Sky/pixel T1, Width T1 \\
    Hazelwood & UT 2015 December 28  & $R$ & 18$\arcmin$ $\times$ 12$\arcmin$ & 0.73$\arcsec$ & 300 & Airmass, X(FITS), Y(FITS) \\
    Myers & UT 2016 January 11 & $B$ &15.5$\arcmin$ $\times$ 15.5$\arcmin$ & 0.92$\arcsec$&300 & Airmass\\
    Myers & UT 2016 February 14 & $I$ &15.5$\arcmin$ $\times$ 15.5$\arcmin$ & 0.92$\arcsec$&30 & Airmass \\
    Harlingten & UT 2016 July 27 & $i^\prime$ & 27$\arcmin$ $\times$ 27$\arcmin$ & 0.8$\arcsec$ & 180 &  None \\
    AAT IRIS2 & UT 2016 October 16 & $Ks$ & 7.7$\arcmin$ $\times$ 7.7$\arcmin$ & 0.45$\arcsec$ & 2 & Time \\
    EABA    &   UT 2016 December 05   &   $I$&  8$\arcmin$ $\times$ 12$\arcmin$&0.25$\arcsec$ & 30 & Airmass     \\
     \hline
    \hline
 \end{tabular}
\begin{flushleft}
  \footnotesize \textbf{\textsc{NOTES}} \\
  \footnotesize All the follow-up photometry presented in this paper is available in machine-readable form in the online journal.
  \end{flushleft}
\end{table*}

\subsection{Spectroscopic Follow-up}\label{sec:spec}
A series of spectroscopic follow-up observations were performed to characterise the atmospheric properties and radial velocity variations of KELT J041621-620046. These observations were performed using the Wide Field Spectrograph (WiFeS) on the ANU 2.3m telescope at Siding Spring Observatory, Australia. WiFeS is an image slicer integral field spectrograph, with a spatial resolution of 1\arcsec per spatial pixel in the $2\times$ bin mode. Our observing strategy, reduction, and analyses techniques are detailed in full in \citet{Bayliss:2013, Zhou:2015}.

\begin{figure}[!ht]
  \centering
  \includegraphics[width=0.99\linewidth]{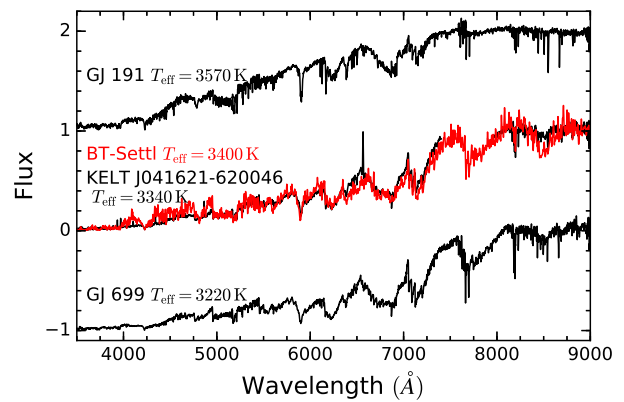}
 \vspace{-0.1in} 
  \caption{WiFeS $R=3000$ spectrum of KELT J041621-620046. The best match BT-Settl model spectrum is over-plotted for comparison. We also plot the spectrum of GJ~191 and GJ~699, two well characterised M-dwarfs, for visual comparison. }
   \label{fig:lowres_spec}
\end{figure}

Spectroscopic classification of the binary was obtained with a $\lambda / \Delta \lambda \equiv R = 3000$ spectrum, covering the wavelength range of 3500-9000\space\AA. The spectrum was flux calibrated as per \citet{Bessell:1999}, using spectrophotometric standard stars eg 131, HD 26297, HD 29574, HD 36702  (with spectrophotometric data from \citealt{Hamuy:1992} and \citealt{Bessell:1999}) observed on the same night.  The flux-calibrated, low-resolution spectrum of KELT J041621-620046 is plotted in Figure \ref{fig:lowres_spec}. While this is a composite spectrum of both stellar components, we match the spectrum to the synthetic spectral templates for a first approximation of the stellar parameters. For the spectral matching, we adopt BT-Settl atmosphere models  \citep{Allard:2012}, with \citet{Asplund:2009} abundances. The surface gravity is fixed to $\log g=5$, as this is the expected gravity for M-dwarfs \citep[e.g.][]{Baraffe:1998,Dotter:2008}.  We find a best fit effective temperature for KELT J041621-620046 of \teff = 3340$\pm$85\,K. The effect of surface gravity on the estimated temperature is small; using models with $\log g = 4.5$ yields a binary temperature only 15\,K lower, insignificant compared to the uncertainties. The WiFeS spectra of GJ 191 \citep{Segransan:2003}, which has a \teff = 3570$\pm$156\,K, and GJ~699 which has a \teff = 3224$\pm$10\,K  \citep{Allard:2012}, are plotted for comparison. Metallicity is estimated by measuring the $\zeta_{TiO/CaH}$ index \citep{Reid:1995} using the calibration from \citet{Lepine:2013}, finding a metallicity of [M/H]=-0.2$\pm$0.2. As described in \citet{Zhou:2015}, we measure log(L$_{H\alpha}$/L$_{bol}$) for KELT J041621-620046A and B to be -3.903 $\pm$ 0.021 and -3.839 $\pm$ 0.050, respectively. Using the relationship between log(L$_X$/L$_{bol}$) and log(L$_{H\alpha}$/L$_{bol}$), as described in \S2.2 of \citet{Stassun:2012}, we estimated log(L$_X$/L$_{bol}$) to be -3.3$\pm$1.1 for KELT J041621-620046A and -3.2$\pm$1.1 for KELT J041621-620046B.

\begin{figure}[!ht]
  \centering\includegraphics[width=0.99\linewidth,trim = 0 0 0 0]{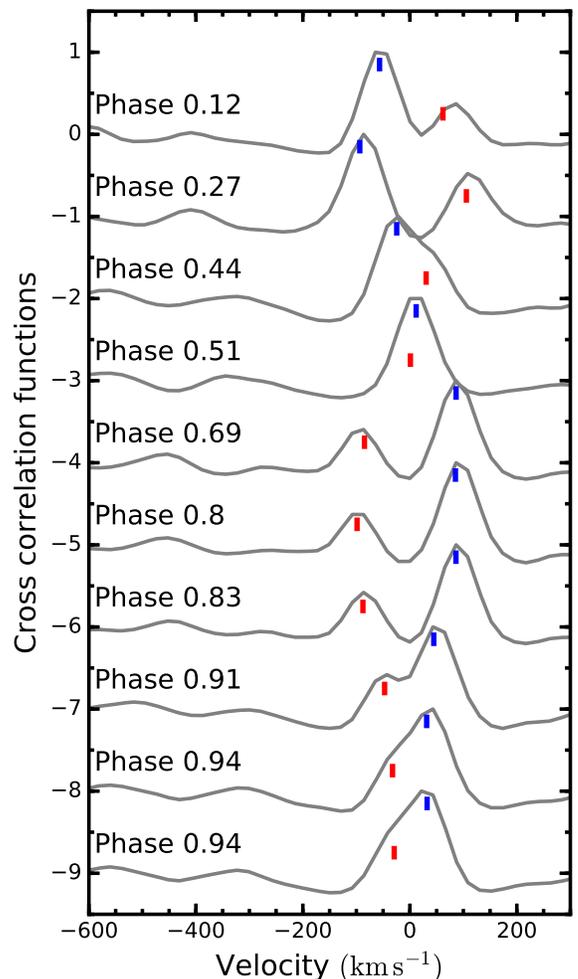}
 \vspace{-0.1in} 
  \caption{An example series of cross correlation functions from WiFeS spectra of KELT J041621-620046, against an M4.0 radial velocity template observed by WiFeS. The derived velocities for each stellar component are marked by the blue and red ticks.}
  \label{fig:CCFs}
\end{figure}

\begin{figure}[!ht]
  \centering\includegraphics[width=0.99\linewidth,trim = 0 0 0 0]{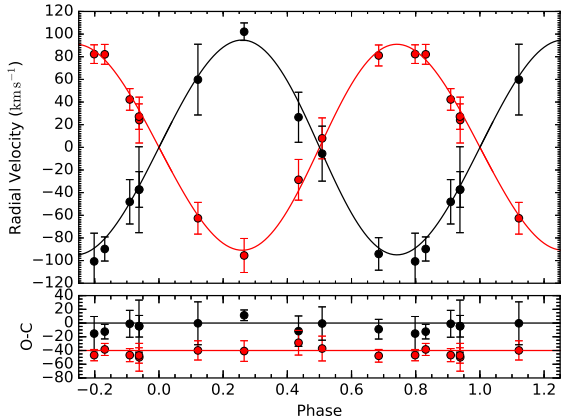}
 \vspace{-0.1in} 
  \caption{Follow-up radial velocities from WiFeS for KELT J041621-620046A (Red) and B (Black). The best-fit models for A and B are shown in black and red, respectively.}
  \label{fig:RVs}
\end{figure}

Radial velocities are measured from WiFeS multi-epoch medium-resolution observations, at $R=7000$, over the wavelength range of $5200-7000$ \AA. A total of ten observations were obtained over UT 2015 September 30 to UT 2015 October 05 (See Table \ref{tab:RVs} and Figure \ref{fig:RVs}). To measure the radial velocities of both stellar components in the spectra, we cross-correlate the spectra against nine M-dwarf standards observed by WiFeS, ranging over the spectral classes of M1.5 to M4.0. An example series of cross correlation functions with an M4.0 template are shown in Figure~\ref{fig:CCFs}. To derive the velocities for both stellar components from each exposure, we simultaneously fit double Gaussians to the CCF from all the exposures that were gathered. The CCF from each exposure are described by the free parameters velocity centroids $v_1$, $v_2$, while the parameters for light ratio $L_2/L_1$, and CCF widths are shared amongst all exposures. The best fit parameters and per-point uncertainties are estimated from a Markov Chain Monte Carlo (MCMC) analysis, using the \emph{emcee} implementation of an affine invariant ensemble sampler \citep{Foreman:2013}. We apply this same velocity fitting procedure to the sets of CCFs derived from each M-dwarf template to understand the template spectral type dependence of the radial velocity measurements. The scatter in the velocity measured for each point for the set of models is then added in quadrature to the mean velocity uncertainty from the MCMC analysis. We also measure a light ratio of $L_2/L_1=0.43\pm0.03$ from the relative heights of the CCFs; this is subsequently used to constrain the global fitting. As an independent check on the reported RVs in Table \ref{tab:RVs}, We also ran 2-D cross-correlations with TODCOR \citep{Zucker:1994} to rule out the possibility that our 1-D cross-correlations may have introduced systematic velocity shifts due to line blending. The TODCOR-derived RV semi-amplitudes are consistent to within 0.25$\sigma$ of those derived via the 1-D cross-correlations. Given this agreement and the quality of the WiFeS spectra, we conclude that the RVs are most strongly limited by potential systematics in the wavelength solutions rather than the analysis technique.

As with many other short period M-M binaries, the Balmer and Calcium $H$ \& $K$ lines are found in emission due to stellar activity \citep{Metcalfe:1996, Zhou:2015}. \citet{Lopez:2007} found a correlation between the activity index and the relative model-observation radius discrepancy of M-dwarf binaries. We use the WiFeS $R=7000$ spectrum to estimate the H$\alpha$ flux in each stellar component of KELT J041621-620046. We measure H$\alpha$ luminosities of $\log L_{H\alpha} / L_{Bol} = -3.7 \pm 0.1$ and $-4.0\pm0.1$ for the two components of KELT J041621-620046, derived from the two WiFeS exposures taken on UT 2015 October 02 and UT 2015 October 05 at phase quadratures.

\begin{table}
 \centering
 \setlength\tabcolsep{1.5pt}
\caption{KELT J041621-620046 RV observations with WiFeS.}
 \begin{tabular}{ccccc}
 \hline
 \hline
  $\bjdtdb$ & RV$_1$ & $\sigma$RV$_1$ & RV$_2$ & $\sigma$RV$_2$ \\
   & (\kms)&  (\kms) & (\kms) & (\kms) \\
 \hline
2457296.21686 & 26.10 & 8.52 & -35.20 & 6.62 \\
2457298.15719 & 83.00 & 3.90 & -92.26 & 6.04\\
2457299.07271 & 9.78 & 7.60 & -3.72 & 10.21\\
2457300.10221 & -26.96 & 7.56 & 28.27 & 9.30\\
2457301.02527 & -93.83 & 6.32 & 103.74 & 3.27\\
2457325.07740 & 41.37 & 4.01 & -49.02 & 8.24\\
2457462.90894 & 27.13 & 4.73 & -37.45 & 15.96\\
2457464.97631 & 82.46 & 3.47 & -100.55 & 10.48\\
2457473.90325 & 83.25 & 3.69 & -88.62 & 4.48\\
2457490.89514 & -59.70 & 5.89 & 62.72 & 13.14\\
\hline
 \hline
\end{tabular}
 \label{tab:RVs}
\begin{flushleft}
\end{flushleft}
\end{table}

\section{\bf{Analysis and Results}}
\label{sec:results}

\subsection{Spectral Energy Distribution Fit}
\begin{figure}[!ht]
 \centering \includegraphics[width=\columnwidth,trim=40 20 20 20,clip]{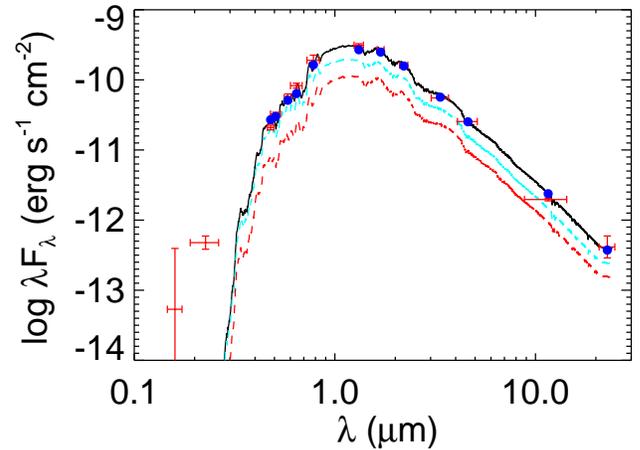}
 \caption{The SED for KELT J041621-620046 where the red points are the photometry and error from the literature.Crosses represent the measured fluxes, with vertical bars representing the measurement uncertainties and the horizontal bars representing the width of the bandpass. The cyan and red curves represent the best fitting NextGen synthetic spectra for the primary and secondary, respectively (see the text). The black curve corresponds to the sum of the primary and secondary model SEDs, and the blue points are the model passband fluxes.}
 \label{fig:SED_figure}
\end{figure}

To estimate the ``average" effective temperature of the stars, we first fit the combined-light spectral energy distribution (SED) of the system using catalog photometry from GALEX, APASS, 2MASS, and WISE spanning a wavelength range of 0.15--20 $\mu$m, as shown in Figure \ref{fig:SED_figure} and listed in Table \ref{tbl:Host_Lit_Props}. The fitted SED model is a NextGen stellar atmosphere model with free parameters of $T_{\rm eff}$ and $A_V$ (we adopted a main-sequence surface gravity of 5.0 and solar metallicity). This initial fit yielded a best-fit $T_{\rm eff} = 3350 \pm 50$ K and $A_V = 0.03 \pm 0.03$ mag, in full agreement with the temperature measured by WiFeS. The SED fitting is only used as a consistency check to spectroscopic analysis and the global fit results.  In Figure \ref{fig:SED_figure}, a ultra-violet excess is clearly seen relative to our SED model, likely coming from the chromosphere and the transition region of the stars.  

\begin{figure*}[!ht]
  \centering\includegraphics[width=0.99\linewidth,trim = 0 0 0 0]{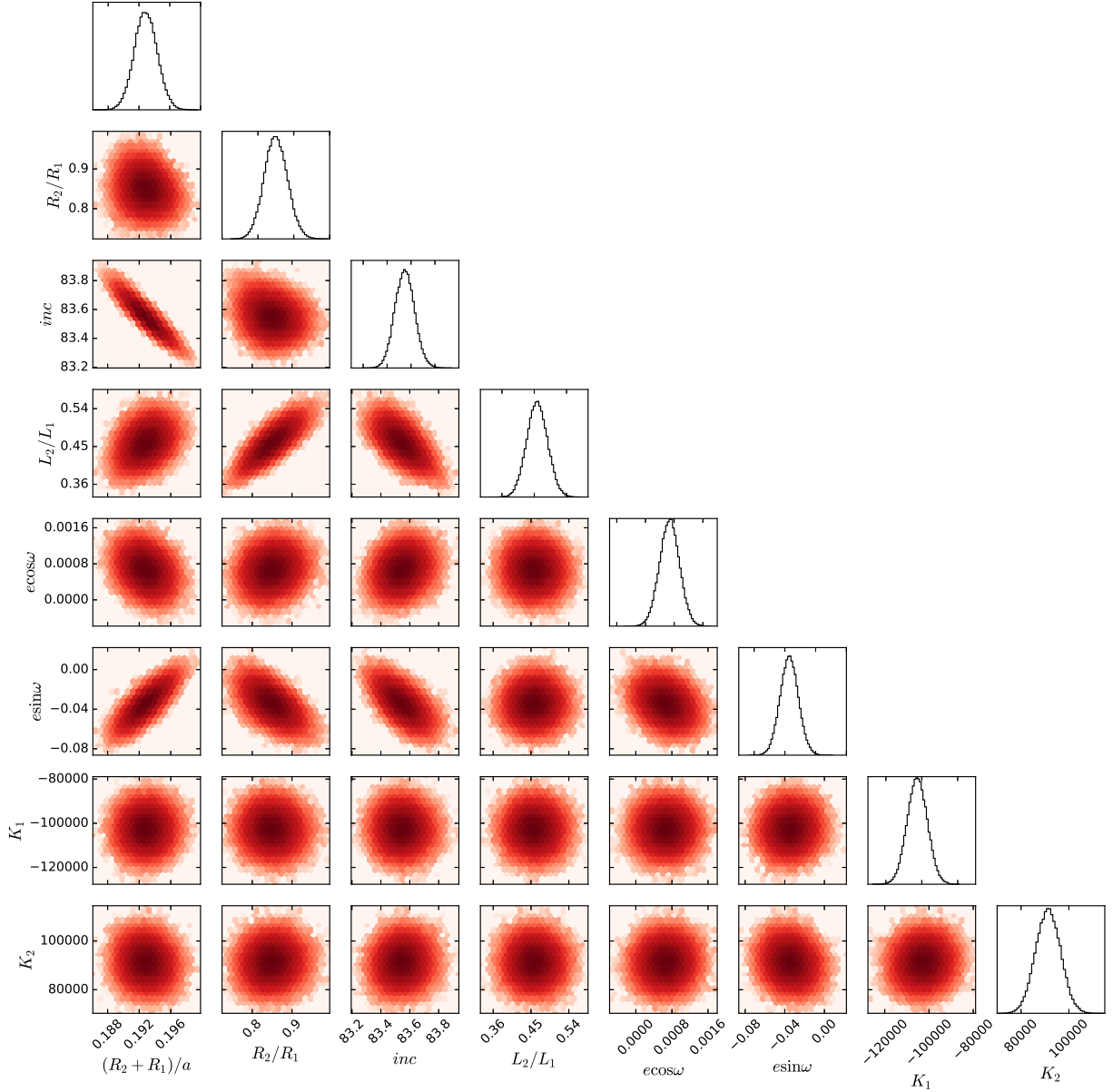}
 \vspace{-0.1in} 
  \caption{Posterior probability distribution of key global fitting free parameters. We note significant covariance in the eclipse modelling parameters, such as inclination $inc$, radius sum $(R_1+R_2)/a$, light ratio $L_2/L_1$, and the eccentricity parameters $e\cos \omega$ and $e\sin \omega$. }
  \label{fig:corner}
\end{figure*}

\subsection{Global Model}
\label{sec:Model}
We perform a global model fit of the follow-up photometry and RVs using EBOP \citep{Nelson:1972,Popper:1981}. Following \citet{Zhou:2015}, the eclipses are modeled using a modified version of the JKTEBOP code \citep{Southworth:2004}, with free parameters of Period $P$, time of eclipse $t_0$, radius ratio $R_2/R_1$, normalized orbital radius $(R_1+R_2)/a$, inclination $i$, light ratio $L_2/L_1$, radial velocity semi-amplitudes $K_1$ and $K_2$, and orbital parameters $e\cos \omega$ and $e\sin \omega$. Limb darkening coefficients for each photometric band are fixed to values interpolated from \citet{Claret:2000} using the Phoenix models. Unfortunately, the follow-up light curves are not precise enough to derive the gravity darkening or reflection coefficients. Therefore, we fix these values to be 0.2 and 0.5, respectively, based on the work of \citet{Morales:2009}. We assign a Gaussian prior on the light ratio based on the height ratio of the CCF peaks in the WiFeS spectra, and assume it to be identical across the bands. Since both stars are of approximately the same spectral type, we assume the light ratio is the same across all light curves. The posterior distribution is derived using an MCMC analysis with the \emph{emcee} package \citep{Foreman:2013}. The results of the EBOP model are presented in Table~\ref{tbl:results} and the posterior probability distribution of key global fitting free parameters is shown in Figure \ref{fig:corner}.

\begin{figure*}[!ht]
\centering\includegraphics[width=0.9\linewidth, trim = 0 3in 0 0,clip]{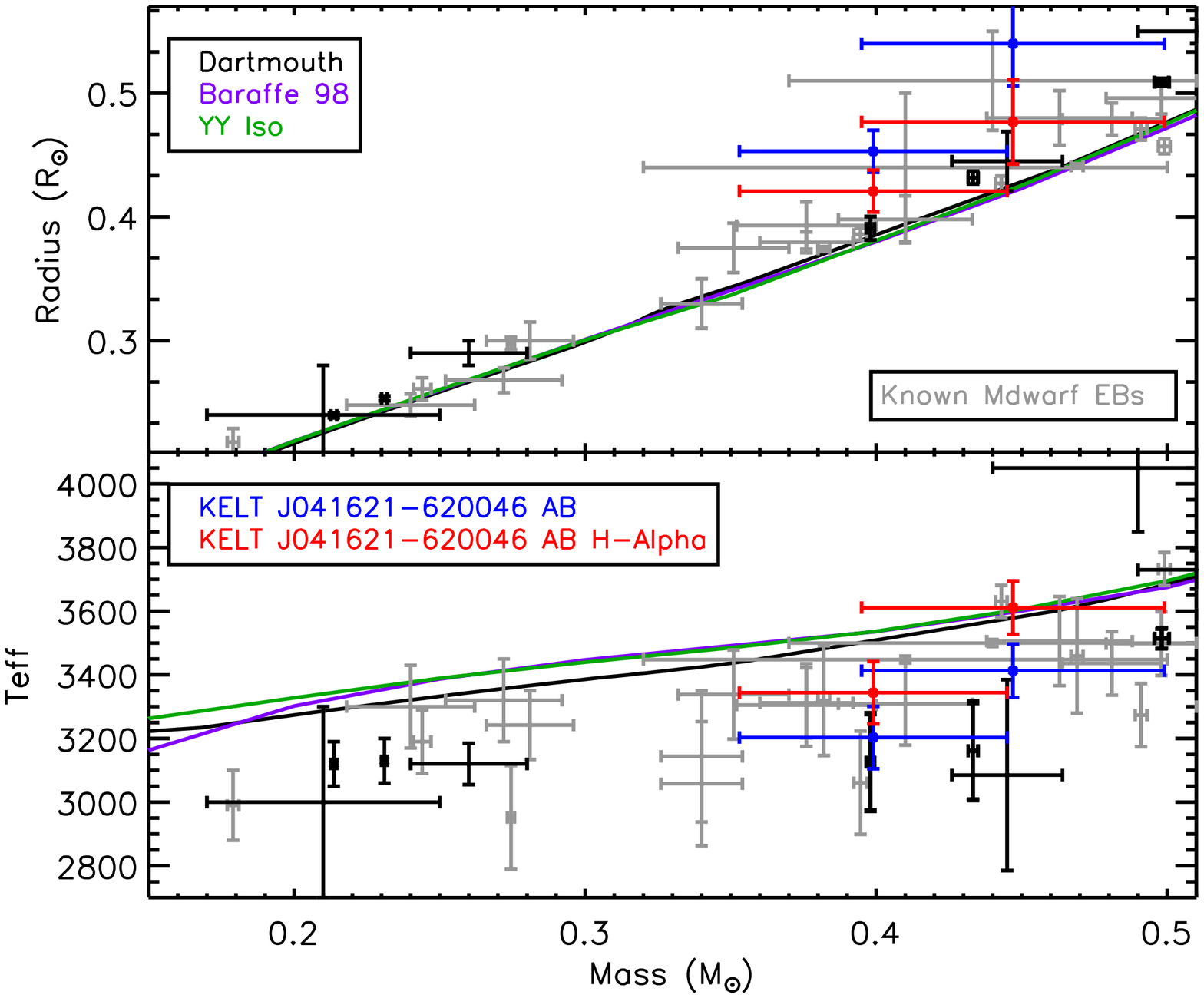}
  \caption{(Top) The measured radii, temperatures, and masses for all known M-dwarfs in double-lined eclipsing binary systems are shown in gray. Systems with $V\ge$14 are shown in black. Also shown are the Dartmouth (black line, \citealp{Dotter:2008}), Baraffe 1998 (purple line, \citealp{Baraffe:1998}), and Yonsei-Yale (green line, \citealp{Yi:2001,Spada:2013}) solar metallicity, 5 Gyr models. The KELT J041621-620046 system is shown in blue (without H$\alpha$ correction) and in red (with H$\alpha$ correction applied); see the text. The known M-M binaries plotted were obtained from \citet{Birkby:2012, Nefs:2013, Dittmann:2016}.}
\label{fig:KnownSystems}
\end{figure*}

We can estimate the individual stellar temperatures from the global modeling results. We assume that the luminosity of both stars can be well described by the Stefan-Boltzman's law, and that the system, together, has an `effective' binary temperature of \teff = 3340$\pm$85\,K as measured by WiFeS. We can then adopt the individual stellar radii and luminosity ratio determined from the global modelling, and derive an effective temperature of $3481\pm83$\,K and $3108\pm75$\,K for the primary and secondary stars, respectively. We also check this result against the SED independent of the global fit. We re-fit the combined-light SED as above, but this time using the sum of two stellar atmospheres whose flux-weighted average temperature is 3350 K (from the initial SED fit above) and whose temperature ratio is as given from the light curve modeling. We adopt the spectroscopic light ratio from the WiFeS spectra of $L_2/L_1 = 0.43\pm 0.03$ over the wavelength range 0.52--0.70 $\mu$m. The only free parameter then is the radius ratio required to produce a flux ratio in the 0.52--0.70 $\mu$m range of 0.43$\pm$0.03. This resulting fit shown in Figure \ref{fig:SED_figure} yields individual temperatures of $T_1 = 3413 \pm 84$ K and $T_2 = 3203 \pm 98$ K, and a radius ratio of $R_2/R_1 = 0.838\pm0.029$, consistent with the radius ratio from the global modeling, and the temperature ratio derived from Stefan-Boltzman's law. The final system parameters are summarized in Table~\ref{tbl:results}. In addition, the component stellar masses, radii, and temperatures are shown together in Fig.~\ref{fig:KnownSystems} in comparison to other low-mass EBs from the literature and to theoretical stellar isochrones.


To clarify the system geometry, KELT J041621-620046 A is the primary star due to its higher mass, radius, and luminosity. The primary eclipse of the system is when star B passes in front of A causing the deeper eclipse seen in Figures \ref{fig:DiscoveryLC} and \ref{fig:Followup}. Both the primary and secondary eclipses are partial/grazing eclipses and not transits. All figures except Figure \ref{fig:DiscoveryLC} are using the global fit ephemeris in ($\bjdtdb$), with the primary eclipse a Phase = 0. In addition to the eclipses, our global analysis infers that the light curves should exhibit out-of-transit variations at the level of 2\,mmag peak to peak, due to the primary and secondary stars being oblate at 0.2\% and 0.1\%, respectively. Additionally, we removed the eclipses from the KELT lightcurve and analyzed the out-of-eclipse variability. The results of this analysis suggest that the rotation periods of both stars are synchoronized to the orbital period. However, due to the low precision of the KELT observations for a target this faint, we do not claim spin-orbit synchronization. 


\begin{table*}
 \scriptsize
\centering
\setlength\tabcolsep{1.5pt}
\caption{Median values and 68\% confidence interval for the physical and orbital parameters of the KELT J041621-620046 system}
  \label{tbl:results}
  \begin{tabular}{lccccc}
  \hline
  \hline
  Parameter & Description (Units) &  KELT J041621-620046A & KELT J041621-620046B & KELT J041621-620046A$^\dagger$ & KELT J041621-620046B$^\dagger$  \\
   &  & {\bf All Eclipses} &{\bf All Eclipses} &Selected Eclipses  & Selected Eclipses  \\
   &  & {\bf Adopted} &{\bf Adopted} &  &  \\
 \hline
                               ~~~$M_{*}$\dotfill &Mass (\msun)\dotfill & $0.447^{-0.047}_{+0.052}$ & $0.399^{-0.042}_{+0.046}$& $0.447^{-0.047}_{+0.052}$ & $0.399^{-0.042}_{+0.046}$\\
                             ~~~$R_{*}$\dotfill &Radius (\rsun)\dotfill &$0.540^{-0.032}_{+0.034}$ & $0.453\pm0.017$ &$0.540^{-0.032}_{+0.034}$ & $0.453\pm0.017$ \\
                             ~~~$R_{2}/R_{1}$\dotfill & SED Radius Ratio (Fixed)\dotfill & \multicolumn{4}{c}{\dotfill$0.838\pm0.029$\dotfill} \\
                             ~~~$(R_1+R_2)/a$\dotfill &Normalized radius sum\dotfill & \multicolumn{4}{c}{\dotfill$0.1948\pm0.0014$\dotfill} \\
                             ~~~$L_{2}/L_{1}$\dotfill &Luminosity Ratio\dotfill & \multicolumn{4}{c}{\dotfill$0.462\pm0.026$\dotfill} \\
                  ~~~$P$\dotfill &Period (days)\dotfill & \multicolumn{2}{c}{\dotfill$1.1112860702^{-0.000000376}_{+0.000000379}$\dotfill} & \multicolumn{2}{c}{\dotfill$1.1112862048\pm0.000000052$\dotfill}\\ 
                  ~~~$T_{0}$\dotfill &Time of eclipse ($\bjdtdb$)\dotfill & \multicolumn{2}{c}{\dotfill$2455255.96335^{-0.00070}_{+0.00071}$\dotfill}& \multicolumn{2}{c}{\dotfill$2455255.96452\pm 0.00115$\dotfill} \\%
                ~~~$K$\dotfill &RV semi-amplitude (km/s)\dotfill & $91.3\pm4.7$  & $102.5\pm{5.1}$& $90.7\pm4.6$  & $102.2\pm{5.1}$ \\
         ~~~$\teff$\dotfill & Effective temperature (K) \dotfill & 3481 $\pm$ 83 & 3108$\pm$75 & 3462 $\pm$ 82 & 3125$\pm$76\\ 
                          ~~~$i$\dotfill &Inclination (degrees)\dotfill & \multicolumn{2}{c}{$83.4\pm0.1$}& \multicolumn{2}{c}{$83.6\pm0.1$} \\
        ~~~$\ecosw$\dotfill & \dotfill & \multicolumn{2}{c}{\dotfill$0.0007\pm0.0002$\dotfill} & \multicolumn{2}{c}{\dotfill$0.0005\pm0.0003$\dotfill}\\
        ~~~$\esinw$\dotfill & \dotfill & \multicolumn{2}{c}{\dotfill$-0.034\pm0.011$\dotfill} & \multicolumn{2}{c}{\dotfill$-0.021\pm0.011$\dotfill}\\
        ~~~$u_{1\,B}$\dotfill & $B$ band linear limb darkening coefficient\dotfill & 0.3746 & 0.5463 & --- & ---  \\
        ~~~$u_{2\,B}$\dotfill & $B$ band quadratic limb darkening coefficient\dotfill & 0.5043 & 0.3790 & --- & --- \\
        ~~~$u_{1\,V}$\dotfill & $V$ band linear limb darkening coefficient\dotfill & 0.4335 & 0.6038 & --- & --- \\
        ~~~$u_{2\,V}$\dotfill & $V$ band quadratic limb darkening coefficient\dotfill & 0.4384 & 0.3229 & --- & --- \\
        ~~~$u_{1\,R}$\dotfill & $R$ band linear limb darkening coefficient\dotfill & 0.3709 & 0.5074 & --- & --- \\
        ~~~$u_{2\,R}$\dotfill & $R$ band quadratic limb darkening coefficient\dotfill & 0.4554 & 0.3697 & --- & --- \\
        ~~~$u_{1\,I}$\dotfill & $I$ band linear limb darkening coefficient\dotfill & 0.1467 & 0.2815 & --- & --- \\
        ~~~$u_{2\,I}$\dotfill & $I$ band quadratic limb darkening coefficient\dotfill & 0.6311 & 0.5669 & --- & --- \\
        ~~~$u_{1\,Ks}$\dotfill & $Ks$ band linear limb darkening coefficient\dotfill & -0.1077 & -0.1066 & --- & --- \\
        ~~~$u_{2\,Ks}$\dotfill & $Ks$ band quadratic limb darkening coefficient\dotfill & 0.5721 & 0.5217 & --- & --- \\
\hline
 \hline
 \end{tabular}
\begin{flushleft}
  \vspace{-.2in}
 \end{flushleft}
 \footnotesize \textbf{\textsc{NOTES}}\\
 \footnotesize $^\dagger$ The asymmetric Myers I band observations from UT 2016 February 14 was excluded.\\
 \footnotesize The determined limb darkening coefficients were the same for both fits. \\
  \vspace{.1in}
\end{table*}

The Myers I band primary eclipse light curve on UT 2016 February 14 appears somewhat asymmetric, potentially due to a spot-crossing event. Responding to a valuable initiative proposed by our manuscript referee, we omitted the light curve and refit the entire dataset, obtaining $R_1 = 0.555\pm0.033\,M_\odot$, $R_2 = 0.452\pm0.017\,M_\odot$, consistent with our results when including the Myers I band observations to within uncertainties. The results from both fits are shown in in Table \ref{tbl:results}. We adopt the system parameters determined from the global fit that includes all observations. In addition, our radial velocity measurements were obtained by cross correlating the WiFeS observations against a series of M-dwarf standard stars with spectral classes ranging from M1.5 to M4.0. The spectral mismatch between KELT J041621-620046 and the standard stars contributed to the relatively large per-point velocity uncertainties. To test the impact of mismatched templates on our radial velocity orbit solution, we re-derived the velocities using only the M3.4 and M4.0 standard stars, deriving $M_1 = 0.45 \pm 0.05\,M_\odot$ and $M_2 = 0.40\pm0.04\,M_\odot$, within $1\sigma$ of the results from Table~\ref{tbl:results}.

\section{\bf{Discussion}}
\label{sec:Discussion}

KELT J041621-620046 joins the ranks of a small number of double-lined eclipsing binary systems in which both stellar components are M-dwarfs (Fig.~\ref{fig:KnownSystems}). 
primary and secondary components have masses of $\approx$0.45 and $\approx$0.40\msun, respectively, and thus occupy an interesting region of parameter space at or near the fully convective boundary. 

In addition, as with many of the other known M-dwarf EBs, the stellar radii and effective temperatures differ significantly from the predictions of standard theoretical stellar isochrone models. As shown in Fig.~\ref{fig:KnownSystems} (blue symbols), both stars appear to have a radius inflated by 17--28\% and an effective temperature suppressed relative to the same solar metallicity stellar models by 4--10\%; the latter exceeds the 2.2\% suppression seen in single M dwarfs by \citet{Mann:2015}, but this effective temperature offset between single and binary M dwarfs is known \citep{Boyajian:2012}. The stellar models shown in Fig.~\ref{fig:KnownSystems} only change by a few percent when changing solar metallicity by 0.5 dex. This small difference is within our reported errors shown in Table \ref{tbl:results}. KELT J041621-620046B is more consistent with the stellar models in its radius while KELT J041621-620046A has a more consistent temperature. Interestingly, the mass of the two stars in KELT J041621-620046 (M$_A$ = $0.447^{-0.047}_{+0.052}$ \msun and M$_B$ = $0.399^{-0.042}_{+0.046}$ \msun) are similar to those of the CU Cnc system (CU Cnc A = 0.433\msun and B = 0.3980). The orbital period for CU Cnc is 2.77 days, similar to the 1.11 day period we find for KELT J041621-620046 \citep{Ribas:2003}. However, the radii for CU Cnc A and B are 0.432\rsun and 0.391\rsun \citep{Ribas:2003}, significantly smaller than what we measure for KELT J041621-620046 (R$_A$ = $0.540^{-0.032}_{+0.034}$ \rsun and R$_B$ = $0.453\pm0.017$ \rsun) . It is possible that the slightly shorter orbital period of KELT J041621-620046 may be related to the larger observed radii but a comparative study of both systems could shed light on this discrepancy.

\begin{figure}
    \centering
\includegraphics[width=\linewidth, trim = 0 0.3in 0 0]{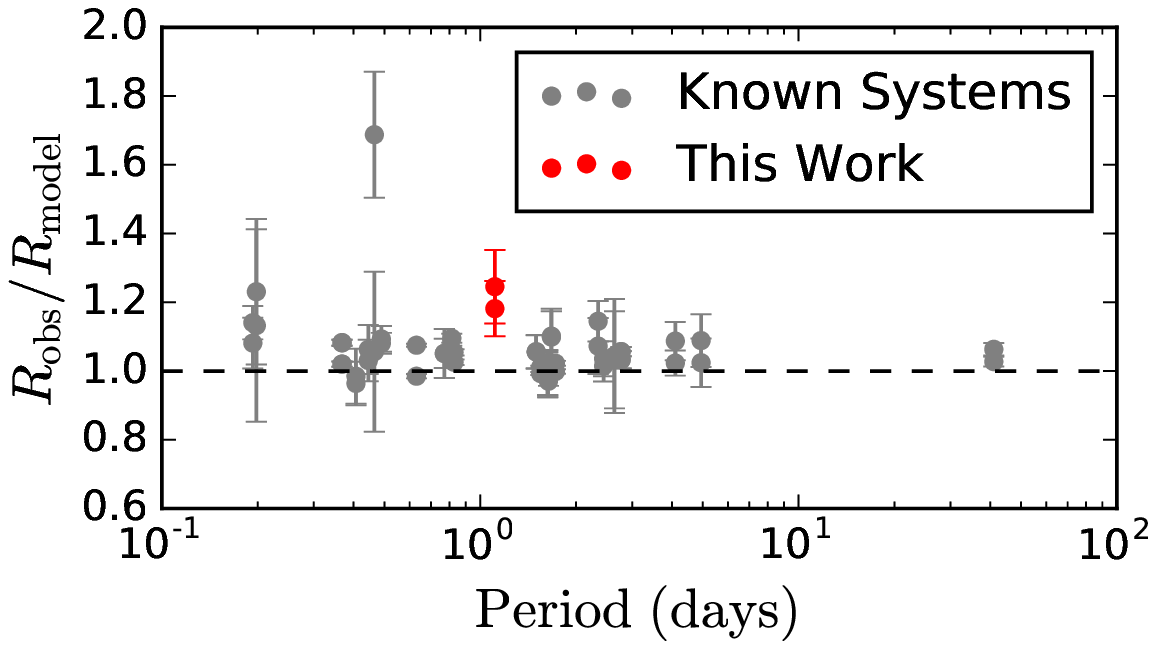}   
\includegraphics[width=\linewidth, trim = 0 0.3in 0 0]{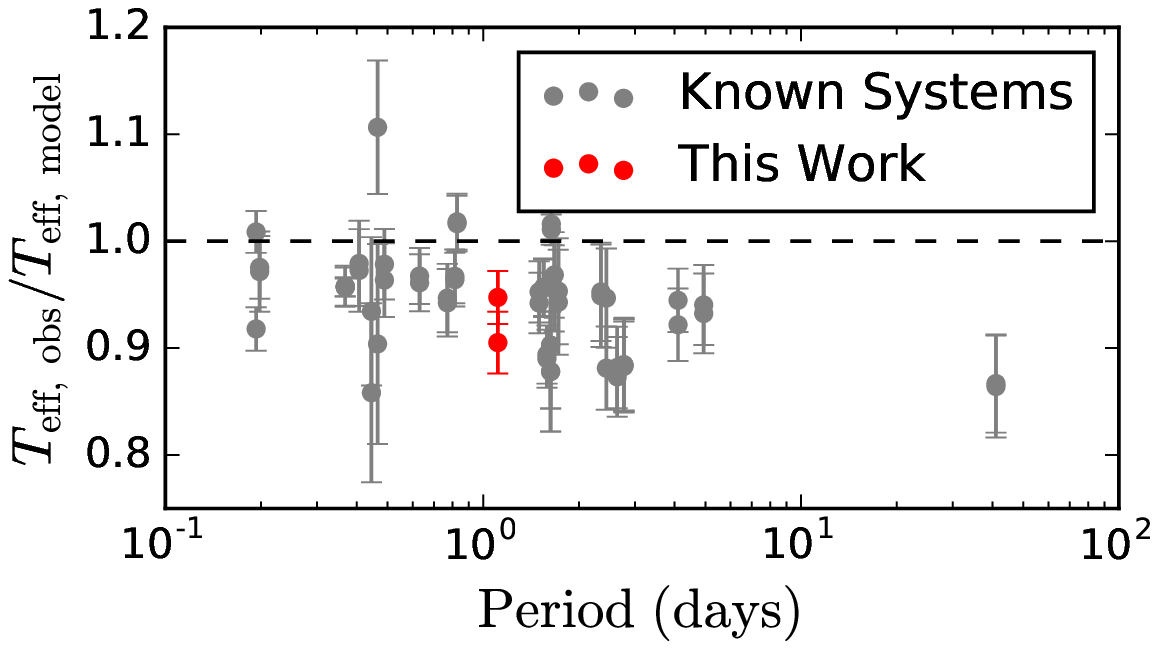}
    \caption{The measured radius relative to the model-predicted radius (\emph{top}) and the measured effective temperature relative to the Baraffe 98 model value (\emph{bottom}) as a function of orbital period for the M dwarfs in Fig. \ref{fig:KnownSystems}.}
    \label{fig:inflationper}
\end{figure}

Fig. \ref{fig:inflationper} shows the difference between observed and Baraffe 1998 model radii and effective temperatures as a function of orbital period for known M-dwarf EBs. While there is no statistically significant correlation observed between radius inflation and orbital period, there is a clear trend towards less temperature suppression for M dwarfs on shorter-period orbits (Spearman rank coefficient $\rho = -0.54; p = 1 \times 10^{-5}$). This observed trend suggests that close binary interactions can dampen the effective temperature suppression. The discovery and characterization of additional M dwarfs in long-period binaries beyond $\sim$40 days would elucidate this trend.

A number of recent studies have demonstrated that chromospheric activity in LMSs can alter their physical properties relative to the expectations of non-magnetic stellar models. In particular, strong activity appears to be able to inflate the stellar radius and to decrease the effective temperature \citep[e.g.,][]{Lopez:2007,Morales:2010}.
Typical amounts of radius inflation and temperature suppression are $\sim$10\% and $\sim$5\%, respectively \citep[e.g.,][]{Lopez:2007}, similar to what we observe here for KELT J041621-620046.

\citet{Stassun:2012} developed empirical relations for the radius inflation and temperature suppression for a given amount of chromospheric H$\alpha$ luminosity. These relations predict that the temperature suppression and radius inflation roughly preserve the bolometric luminosity. The relations are able to explain the surprising reversal of temperatures with mass in the young brown-dwarf eclipsing binary system 2M0535$-$05 \citep{Stassun:2006,Stassun:2007}, including the anomalously cool spectral type of the more massive brown dwarf in the system \citep{Mohanty:2010,Mohanty:2012}, and may also explain some of the observed discrepancies between young stellar eclipsing binaries and non-magnetic stellar models \citep{Stassun:2014b}.

Here, we apply the empirical relations of \citet{Stassun:2012} to KELT J041621-620046 using the observed H$\alpha$ emission of the two components in the system (see Sec.~\ref{sec:spec}). The resulting radii and temperatures (Fig.~\ref{fig:KnownSystems}, red symbols) are brought into better agreement with the predictions of the stellar evolution models for both stars. The agreement seen for both components of KELT J041621-620046 is best with respect to the Dartmouth models, though the agreement with the other models shown is only marginally worse, and the agreement with the models is brought to within $\sim$2$\sigma$. 

To be clear, these H$\alpha$-based adjustments to the stellar radii and temperatures are not ``corrections" per se; the observed radii are in fact inflated and temperatures are in fact suppressed. The adjustments serve to show what the stellar radii and temperatures {\it would be in the absence of magnetic activity}. Evidently, were it not for the strong magnetic activity (as manifested by the strong H$\alpha$ emission) in these rapidly rotating M-dwarfs, their basic properties would be much more in line with the predictions of standard (non-magnetic) stellar models.

\section{\bf{Conclusion}}
\label{sec:Conclusion}

We present the discovery of KELT J041621-620046 as a double-lined eclipsing binary in the field, in which both components are low-mass M-dwarf stars. With component masses of $0.447^{-0.047}_{+0.052}$ and $0.399^{-0.042}_{+0.046}$\msun, and radii of $0.540^{-0.032}_{+0.034}$ and $0.453\pm0.017$\rsun, KELT J041621-620046 becomes one of only a handful of M-dwarfs in eclipsing binaries with precisely determined stellar masses and radii. In addition, the measured stellar masses place the stars at or near the fully convective boundary for M-dwarfs, a particularly important region of stellar parameter space for understanding stellar structure, evolution, and magnetic field generation. 

Both stars appear to be very magnetically active based on their strong H$\alpha$ and Ca II H and K emission. This is perhaps not surprising considering the likely tidal synchronization of the stars with the short-period orbit. Perhaps as a direct consequence of this magnetic activity, KELT J041621-620046A and B appear to have radii that are significantly larger, and effective temperatures that are significantly cooler compared to predictions by standard (non-magnetic) stellar isochrone models. Recent empirical relations for the amount of radius inflation and temperature suppression as a function of chromospheric activity appear able to explain the observed properties of KELT J041621-620046A and the radius of KELT J041621-620046B, and these stars would be in better agreement with theoretical non-magnetic stellar models were they not magnetically active. However, the temperature of KELT J041621-620046B is too low with respect to theoretical non-magnetic stellar models to be explained by magnetic temperature suppression alone. Being quite bright for an M-dwarf eclipsing binary system ($V$ = 13.9), and therefore amenable to more high precision radial velocity follow-up observations, KELT J041621-620046 promises to serve as a testbed for stellar structure and evolution at the stellar fully convective boundary.

\acknowledgments 
J.L.\ acknowledges support from an NSF REU site grant to Vanderbilt University (PHY-1263045). J.E.R.\ and K.G.S.\ acknowledge partial support from NSF PAARE grant AST-1358862. Work performed by J.E.R. was supported by the Harvard Future Faculty Leaders Postdoctoral fellowship. Work by B.S.G. and D.J.S was partially supported by NSF CAREER Grant AST-1056524. This work is partially based on observations obtained with the 1.54-m telescope at Estaci\'on Astrof\'isica de Bosque Alegre dependent on the National University of C\'ordoba, Argentina.

\bibliographystyle{apj}
\bibliography{MM_Binary_Paper}

\end{document}